# Can doping graphite trigger room temperature superconductivity? Evidence for granular high-temperature superconductivity in water-treated graphite powder

By *T. Scheike, W. Böhlmann, P. Esquinazi\*, J. Barzola-Quiquia, A. Ballestar, A. Setzer*

[\*]     E-mail: esquin@physik.uni-leipzig.de

Division of Superconductivity and Magnetism
Institut für Experimentelle Physik II
Fakultät für Physik und Geowissenschaften der Universität Leipzig
Linnéstrasse 5, D-04103 Leipzig, Germany

The existence of room-temperature superconductivity (RTS) has been claimed in 1974 through the observation of Josephson tunneling behavior of the electric current vs. magnetic field applied on aluminum-carbon-aluminum sandwiches.[1] In that work the carbon layer consisted of small, strongly distorted graphite crystallites. Twenty six years later, magnetization measurements suggested also the existence of RTS in parts of oriented graphite samples.[2] Similar claims were published in, for example, n-type diamond surface[3] as well as in palladium hydride[4] later on. None of those reported results, however, were later independently verified such that the existence of RTS remains an apparently unreachable dream in science. The possibility to have high-temperature superconductivity in graphite,[5] at its interfaces,[6] as well as in disordered carbon[7] has been the subject of only a few experimental studies but of a large number of theoretical works in the last 10 years. For example, one expects to find this phenomenon either by doping graphene layers,[8-11] at the graphite surface region due to a topologically protected flat band[12] or in disordered graphite regions.[13] *p*-wave[13] as well as *d*-wave superconductivity[14,11] were predicted. Trying to dope graphite flakes we found that the magnetization of pure, several tens of micrometers grain size graphite **powder and after a simple treatment with pure water shows clear and reproducible granular superconducting behavior with a critical temperature above 300K. The observed magnetic characteristics as a function of temperature, magnetic field and time, provide evidence for weakly coupled grains through Josephson interaction, revealing the existence of superconducting vortices.**

For the preparation of the water-treated graphite powder we mix 100 mg of ultra pure graphite powder (see Experimental details below) into 20 ml distilled water and this mixture is continuously stirred at room temperature. At the beginning of the preparation the graphite grains are swimming on the surface of the water because of their highly hydrophobic property. After ~1 hour one can observe that the graphite powder forms a suspension within the water, which is well homogenous ~1 hour later. After further stirring for 22 h the obtained powder is recovered by filtration with a clean non-metallic filter and we dry it at 100 °C overnight. We have repeated the same procedure several times and checked for the reproducibility of the measured properties. Without exceptions, all prepared samples with the same powder showed the superconducting behavior we present below as long as the powder is not pressed into compact pellets.

The first high-temperature superconducting (HTSC) oxide samples showed typical signs for granular superconductivity, an issue very well described in the literature of the late 80's and in the 90's. In a granular superconductor in which a minor part of the sample shows superconductivity, the magnetic field-induced hysteresis (in field ($H$) and temperature ($T$)) is the tool one uses to provide evidence for superconductivity and the existence of Abrikosov and/or Josephson vortices (fluxons). A method that it does not need any background subtraction and it is rather free from artifacts from the superconducting quantum interferometer device (SQUID)[15] is the measurement of the hysteresis between the magnetic moment $m(T)$ in the zero-field cooled (ZFC) and field cooled (FC) states.

We present first the results for the untreated graphite powder. In Fig. 1(a) we show the total diamagnetic signal at 300 K (right y-axis) and the small hysteresis loop obtained after subtraction of the linear diamagnetic background. This field hysteresis loop of the magnetization with a remanence $M(H=0) \sim 8 \times 10^{-6}$ emu/g after sweeping to the maximum field $H_{max} = 400$ Oe is about one order of magnitude smaller than the one obtained for the water treated powder at the same experimental conditions (see Fig.4 below). Figure 1(b) shows the temperature dependence of the magnetic moment $m(T)$ of the same untreated powder in the ZFC and FC states at $H = 500$ Oe. There is a negligible temperature hysteresis (within a reproducibility of 0.3 µemu) in most of the temperature range, see Fig. 1(c). We will see below that the water treatment enhances drastically both hysteresis beyond experimental errors.

Figure 2(a) shows $m(T)$ for the water-treated powder in the ZFC and FC states at $H = 500$ Oe. The hysteresis is clearly observable and indicates the presence of pinned magnetic entities in the sample. Since magnetic order can be triggered in graphite by introducing defects (or hydrogen[16]) or after chemical treatment,[17] it is necessary to clarify whether the observed behavior is due or not to a ferromagnetic signal. To answer this we have done the same kind of measurement but at different applied fields. The results of the difference between the magnetic moment in the ZFC and FC states, $\Delta m(T) = m_{FC}(T) - m_{ZFC}(T)$ and at different applied fields are shown in Fig.2(b). At applied fields $H \leq 500$ Oe, $\Delta m(T)$ increases with field at all temperatures. However, at the applied fields $H = 10^3$ Oe and $3 \times 10^3$ Oe, $\Delta m(T)$ is nearly the same showing a weak $T$-dependence with a crossing point $T_i$ from small negative to positive values below ~120K, see Fig. 2(b). At $H \geq 5 \times 10^3$ Oe, $\Delta m(T)$ increases again with field at $T <$ $T_i$. We note that $T_i$ shifts to higher T the higher the applied field.[15] The observed behavior is neither compatible with the pinning of magnetic domain walls in general nor with the ferromagnetic behavior of graphite powder in particular,[17] but with granular superconductivity.

We assume that part of the powder is superconducting and that under a magnetic field a superconducting loop can be produced through the grains, which are coupled by intergrain Josephson links. The low field response of this system can be understood by the single junction superconducting quantum interferometer model.[18-21] Two Josephson critical fields can be defined,[18] namely $h_{c1}^J$, i.e. the lower Josephson critical field as the characteristic field above which fluxons penetrate into the sample, and $h_{c2}^J$, the Josephson decoupling field above which the superconducting grains response is that of isolated superconducting grains. At a maximum applied field $H_{max} < h_{c1}^J$ no hysteresis or remanence is observed and the magnetization is given by the macroscopic shielding currents (running through the weakly coupled grains) and the microscopic current loops (circulating around the grains) within their corresponding penetration depths. This non-hysteretic, reversible region can be recognized by measuring the remanent loop width at zero field $\Delta m(H=0) = m_{H+}(H=0) - m_{H-}(H=0)$ as a function of $H_{max}$ at different constant temperatures, see Fig.3. Here $m_{H+,-}$ are the remanent magnetic moments after returning to the $H=0$ state from $+H_{max}$ or $-H_{max}$. We note that the overall behavior of $\Delta m(H=0, H_{max}, T)$ is very similar to that observed in granular[19-21] as well as in single crystalline[22] HTSC oxides. From these results and as described in Fig. 3 we determine $h_{c1}^J(T)$. The obtained $h_{c1}^J(T)$ values are not only one to two orders of magnitude larger than those obtained in granular HTSC oxides[19-21] but they follow an unusual $-\log T$ dependence, see inset in Fig. 3, with an apparent critical temperature $T_c > 10^3$ K defined at $h_{c1}^J(T_c) \rightarrow 0$ Oe after a linear extrapolation. At $h_{c1}^J < H_{max} < h_{c2}^J$, intergranular and London currents around the grains can trap fluxons as well as intragranular Abrikosov vortices (produced by the displacement of the first ones[23]) in an irreversible manner producing hysteresis in the temperature loops, as shown in Fig.2, as well as field hysteresis and a finite loop width $\Delta m(H=0)$, see Fig.4.

Figure 4 shows the field hysteresis loops after subtraction of the linear diamagnetic background (for a given temperature) measured at different $H_{max}$ at 5K (a) and 300K (b). For $h_{c1}^J(T) < H_{max} < h_{c2}^J(T)$ and as expected[18,19] the overall shape of the hysteresis loops resembles basically the one obtained by the Bean model for bulk superconductors.[24] As example, we show in Fig. 4(a) the loop at 5K and $H_{max}$ = 400 Oe with the one calculated with the Bean model and a full penetration field $H_p$ = 270 Oe. At high enough $H_{max}$, however, a change of the slope of the virgin curve $m(H)$ and an anomalous shrinking of the hysteresis width (instead of a saturation) are measured, see Figs.4(a,b). This behavior has been observed also in granular HTSC oxides and it is well understood assuming that at $H_{max} > h_{c2}^J(T)$, but smaller than the lower intragrain critical field $H_{c1}$, there is a transition to a non-hysteretic mode and the magnetic moment in this field range is due to the flux expulsion from the isolated superconducting grains.[19-20] Defining $h_{c2}^J(T)$ at the beginning of the linear reversible region at high fields we obtained the values shown in the inset of Fig. 3. The so obtained $h_{c2}^J(T)$ shows a similar $T$-dependence as $h_{c1}^J(T)$ leading to a constant ratio in the measured temperature range $h_{c2}^J(T)/h_{c1}^J(T) = 4 \pm 0.5$, similar to values obtained in granular HTSC and in agreement with model predictions in terms of weakly coupled superconducting grains.[20] According to this model the here obtained values of the Josephson critical fields suggest much larger Josephson critical current for similar loop parameters than in the HTSC.

A rough estimate of the relative superconducting mass in our samples gives ~ 100 ppm, taking the diamagnetic slope at low fields without any demagnetization factor. Although small, this superconducting yield is of relevance and, interestingly, of the same order as the ferromagnetic one triggered in graphite by chemical methods.[17] With this mass we can estimate the superconducting magnetization and through this, some parameters of the superconducting granular system. From the total magnetization width obtained at $H = 0$ Oe and at $T = 5$ K ($H_{max} = 400$Oe), following the Bean model we estimate a critical intergranular Josephson current density $10^{11}$ A/m$^2$ > $J_c$ ~ $\Delta M/a$ > $10^{10}$ A/m$^2$ for 1µm < $a$ < 10 µm, where $a$ is the corresponding effective grain length. For the same range of $a$ and with the lower Josephson critical field, see Fig. 3, $h_{c1}^J(5K)$ ~ 150 Oe ~ $J_c \lambda_J$ we get a Josephson penetration depth 0.1µm < $\lambda_J(5K)$ < 1 µm, i.e. an order of magnitude smaller than the corresponding length $a$.

The non monotonic behavior with field obtained for $\Delta m(T) = m_{FC}(T) - m_{ZFC}(T)$ shown in Fig.2(b), is compatible with the description provided above. Some further details are worth to mention. The maximum in $\Delta m(T)$ observed at high temperatures near the turning temperature of 300K at $H \leq 500$ Oe, appears to be related to the so-called peak effect in granular superconductors originated by the change of pinning regime between fluxons and Abrikosov vortices.[23] Above 600 Oe, i.e. above the corresponding $h_{c2}^J(T)$ when the Josephson coupling between the grains is destroyed, this peak vanishes and $\Delta m(T)$ decreases to nearly $T$-independent (small negative) values. $\Delta m(T)$ changes to positive below $T_i$, which occurs when $H < h_{c2}^J(T)$. At applied fields $H > 3000$ Oe, $\Delta m(T)$ increases again suggesting that the pinning of intragranular Abrikosov vortices in the decoupled grains starts to play a role. These results indicate that the effective lower intragrain critical field $H_{c1}$ is also very high. Note that $\Delta m(T)$ increases steadily with field[15] above some threshold suggesting that the pinning increases with field up to the largest measured temperature. This pinning increase with field is in principle commonly seen in superconductors but at fields much lower than the upper critical field $H_{c2}(T)$.

Because of the finite vortex pinning strength we expect to measure a time relaxation of the magnetic moment after reaching a field from a different previously prepared field state in the sample. This well known flux creep phenomenon shows a logarithmic time dependence of the form $m(t)/m(0) = 1 - m_1 \ln(1 + t/\tau_0)$, $m_1 \sim k_B T/U_a$, $U_a$ is an apparent flux creep activation energy and $\tau_0$ a time constant that determines a transient stage before the beginning of the logarithmic relaxation.[25] The time dependence of the magnetic moment shown in Fig.5 was obtained at each temperature after cycling the field to 300 Oe and then decreased to the remanent state ($H = 0$). The observed dependence can be very well described by the above logarithmic time function. From the fits values obtained at high temperatures we obtain $U_a \sim 8000$K. This high activation energy value (at H=0 Oe) would mean that the thermally activated vortex or fluxon diffusion would affect the critical current density mainly at high enough temperatures. The measured temperature dependence of the remanent full hysteresis width $\Delta m(H_{max}=500$ Oe, $H=0)(T) \propto J_c(T,H=0)$ shown in the inset of Fig. 3 agrees with this expectation. We note also that the influence of flux creep is one possible reason for the (small) negative and not strictly zero values of $\Delta m(T) = m_{FC}(T) - m_{ZFC}(T)$ at high enough temperatures and fields shown in Fig.2(b). The time relaxation at much higher fields is strongly reduced suggesting also the increase of $U_a$ with field. As an example, we show in the inset of Fig. 5 the time relaxation for the same sample at a field of $10^4$ Oe coming from a stationary state at a field of $2 \times 10^4$ Oe at 5K and 300 K. The relative change with time at this field drastically decreased to values inherent of the system, i.e. the observed time decay is sample as well as sample-temperature independent, see Fig.5.

The described treatment of the graphite powder with water leads to changes of the magnetic properties as shown above. It is possible that the water molecules form a uniform structured layer at the graphite/water interfaces, as it was reported previously.[26] Increasing the water treatment time the hydrophobicity of the graphite grains is reduced leading to a chemisorption driven by a chemical reaction at the exposed surfaces.[27] Which kind of structure or stoichiometry the superconducting graphite part has and whether the superconductivity appears at graphite interfaces (or sandwiches) produced after the water-treated graphite grains clamp together, remains an issue to be clarified in future studies. It may be that the water treatment dopes parts of the grain surfaces with hydrogen and this element may play an important role as has been also observed for the magnetic order found in graphite.[16,17] To check this we have exposed the virgin graphite powder to hydrogen plasma for 75 minutes at room temperature. The prepared powder shows the same characteristics as the water treated one[15] indicating that hydrogen may play a role in this phenomenon. Whether water, hydrogen or oxygen molecules can enter into the graphite lattice and trigger superconductivity, similar to the apparent intercalation processes and superconductivity found in $FeTe_{1-x}S_x$ samples after their treatments with hot alcoholic beverages[28] remains an interesting question to be answered in future studies. Furthermore, the superconducting signals decrease after annealing the water-treated powder in vacuum.[15]

Coming back to the results in Ref. [1], we believe that those results were indeed correct and that the reported metastability with time and the weak reproducibility of the phenomenon has to do with the state of the interfaces or junctions between the graphite grains as well as with a macroscopic size effect when the effective radius of several coupled grains increases. The larger the size of the macroscopic superconducting loop radius $R$ the smaller the critical current density, i.e. $J_c \propto 1/R$ as observed in granular HTSC.[19] In our water treated powder samples this might be achieved just by pressing them. To check this we have pressed the previously treated and measured graphite powder into pellets with two different pressures. As shown in Fig. 4(c) the width of the superconducting hysteresis loop decreases by ~40% at $H = 0$ after compacting the powder (S1) with a pressure of ~18 MPa. For a pellet prepared with three times larger pressure using the same treated powder (S1) the superconducting like hysteresis loop vanishes completely, see Fig. 4(c). Clearly, the vanishing of the superconducting signals by pressing the treated powder and the low yield obtained by our water treatment strongly limit the use of direct electrical transport measurements of such pellets to check for superconducting signals by transport. Instead, transport measurements in micrometer small grains are needed. Although these measurements were not yet done in water treated grains, it is worth to mention the transport results obtained in micrometer small graphite lamellae with a large density of two-dimensional interfaces.[6] The high carrier density found at these interfaces[29] (due to the influence of hydrogen, lattice defects and/or a different graphite stacking) as well as several experimental hints[6] suggest that granular superconductivity exist at those interfaces. The obtained *I-V* characteristic curves as well as the SQUID results obtained for this kind of samples provide evidence for granular superconductivity above 100K.[15] In this case one may ask whether finite fluxon pinning can exist in a superconducting two dimensional interface. Avoiding here any speculation about the origin of pinning and any possible coupling mechanism necessary to trigger superconductivity at the interfaces, we note that magnetization hysteresis loops have been already observed in Bi bicrystals as well as in $Bi_xSb_{1-x}$ bi- and trycrystals where superconductivity has been proved to exist at interfaces.[30]

The origin of the observed RTS and the overall behavior, especially of the large values and the logarithmic *T*-dependence of the Josephson critical fields and the large critical temperature are clearly unconventional. Taking into account that atomic lattice defects as well as hydrogen can be also a source of magnetic order in the graphite structure, future studies should clarify its correlation to RTS. If there is such correlation, we note that *p*-type superconductivity was predicted for graphite,[13] then it should have a strong influence on the fluxon pinning strength and all the superconducting response depending on it.[31] The observed increase of the irreversible behavior can be ascribed to an increase in vortex pinning with field above a certain threshold. Whether this increase is the usual one because of very high $H_{c2}(T)$ values or it is related to the type of the order parameter should be clarified in the future. Finally, we would like to note that although the superconducting yield with our treatment is small and its phase or stoichiometry remains to be clarified, the overall results indicate that room temperature superconductivity appears to be reachable and that the here used or similar methods may pave the way for a new generation of superconducting devices with unexpected benefits for society.

*Experimental*

*Experimental details*

We used an ultrapure graphite powder RWA/T from SGL Carbon GmbH (Werk Ringsdorff, Germany). The measured impurities by particle induced X-ray emission (PIXE) in µg/g are: Al < 11, P < 3.6, S < 2.2, Cl < 1.4, K < 0.75, Ca < 0.53, Ti < 0.29, V < 0.28, Cr < 0.24, Fe < 0.19, Ni < 0.17, Cu < 0.2 and Zn< 0.31. After the water treatment, the elemental analysis of the powder shows a slight increase in the oxygen concentration from 1.85 at % to 2.15 at%. We expect that the hydrogen concentration also increase after treatment. The powder in the virgin, as received state or after water treatment was packed in a polymer foils of less than 1 mg mass and which magnetic background signal  (measured independently)  remained negligible in all measurements. The pressed powder (pellets) were prepared using a pressure cell avoiding any contact of the powder with metallic surfaces.

The magnetic measurements have been done with a SQUID magnetometer from Quantum Design. Because the critical temperature of the superconducting powder appears to be unreachable with the available techniques, a "zero vortex" virgin state is not really achieved. A nearly zero vortex state can be obtained by cycling the field to zero using the oscillating mode of the SQUID system. Field hysteresis loops at field $H_{max}$ > 1000 Oe are certainly possible but the superconducting solenoid used by the SQUID system shows a hysteresis that influences the measured magnetic moment and must be subtracted from the data in a non simple and transparent way. An example for this high field hysteresis due to the solenoid (or other SQUID artifact) can be seen in Ref. [15]. This is the reason why we restrict ourselves to show here only hysteresis loops measured at lower fields. We note that this artifact is negligible in the ZFC-FC hysteresis measurements.

*Acknowledgements*

This research is supported by the Deutsche Forschungsgemeinschaft under Contract No. DFG ES 86/16-1. A.B. was supported by the project ESF-Nano under the Graduate School of Natural Sciences "BuildMona".

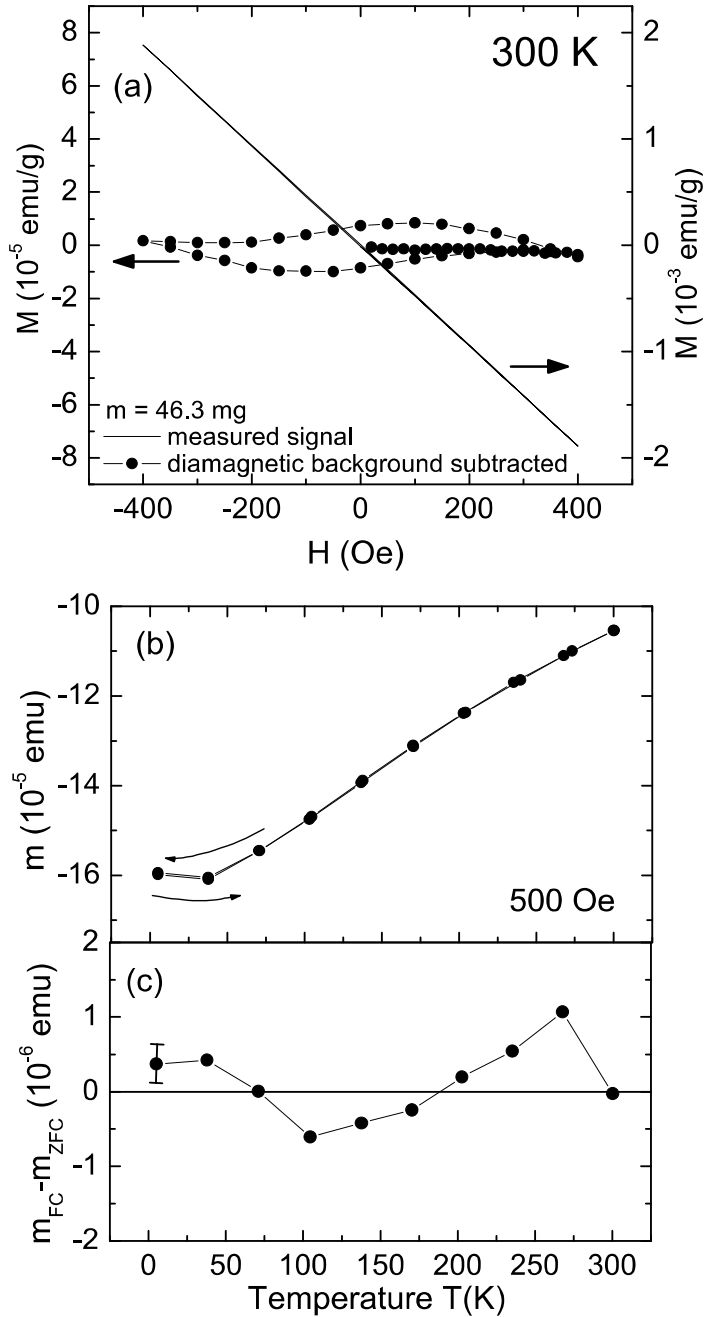

**Figure 1.** (a) Magnetic moment at 300 K measured for the untreated (non-pressed) graphite powder (right y-axis) vs. applied field. Left y-axis: The same data after subtraction of the diamagnetic linear background. (b) Magnetic moment without any subtraction as a function of temperature in the ZFC (lower curve) and FC states at 500 Oe of the same untreated graphite powder. (c) Temperature dependence of the difference between the upper FC and ZFC curves at 500 Oe applied field.

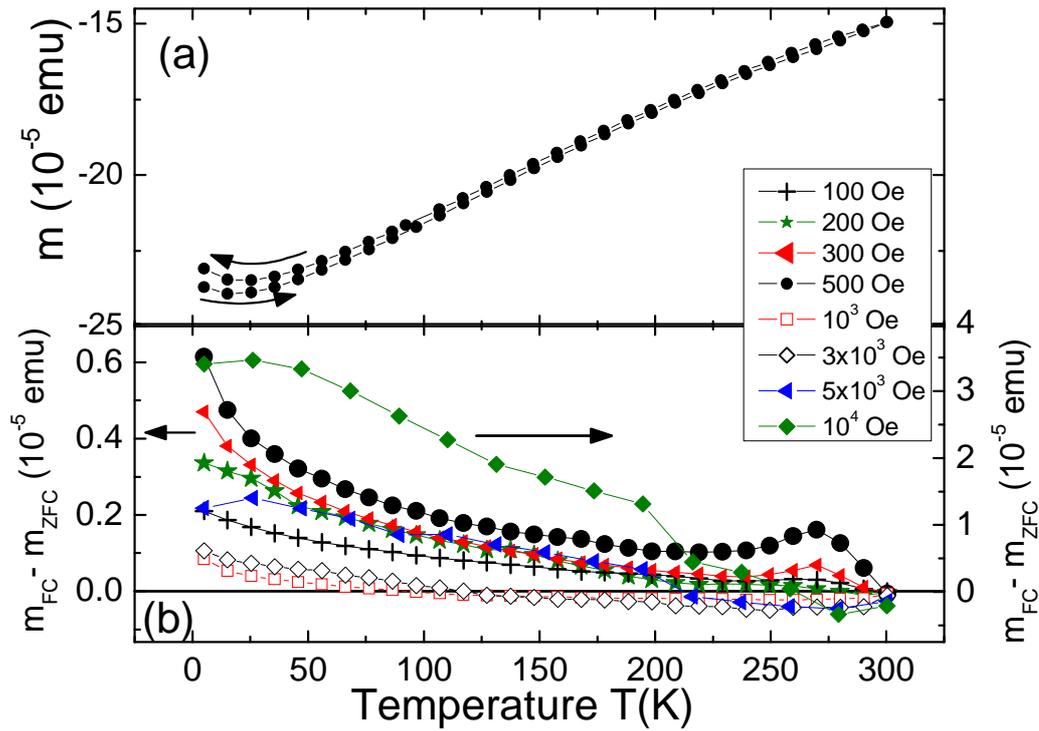

**Figure 2.** (a) Magnetic moment, without any subtraction, as a function of temperature in the ZFC (lower curve) and FC states at 500 Oe of the water-treated graphite powder. (b) Temperature dependence of the difference between the upper FC and ZFC curves at different applied fields. Left y-axis corresponds to the applied fields up to 500 Oe. Right y-axis corresponds to higher applied fields. The results were obtained for a non-pressed water treated graphite powder sample of mass m = 61.5 mg.

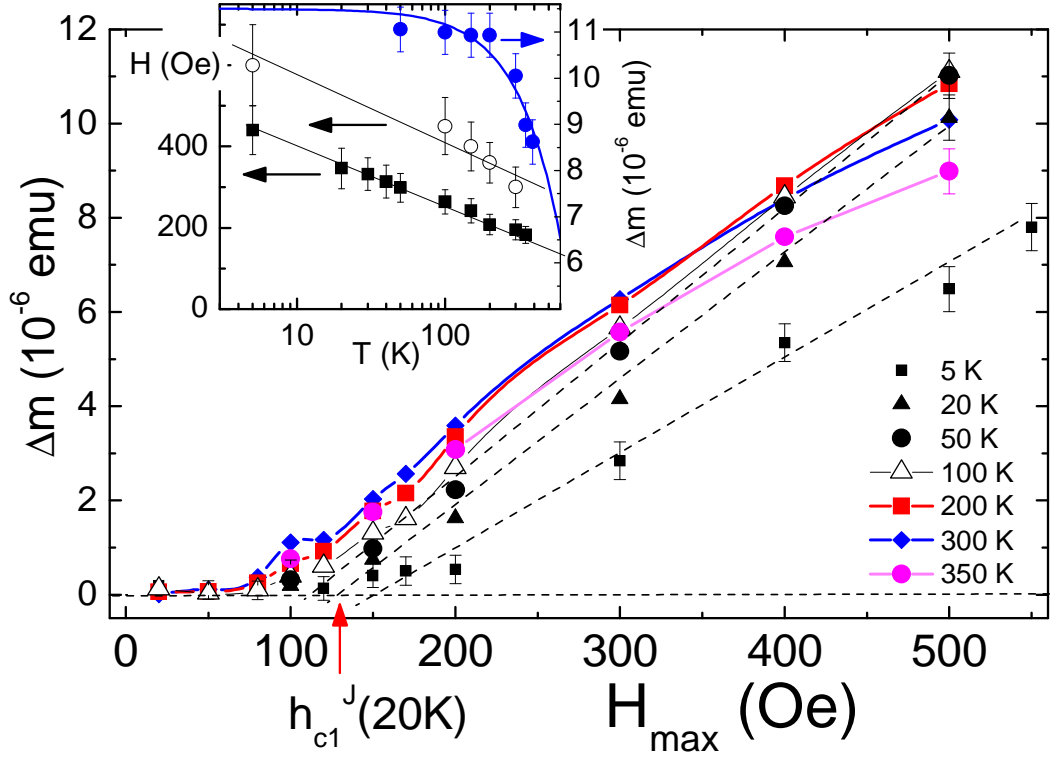

**Figure 3.** Full remanent (H=0) magnetic moment width Δm of the hysteresis loops measured after cycling the sample (61.5 mg) to a maximum field $H_{max}$ at different temperatures. The lower Josephson critical field $h_{c1}^J$ is defined at the crossing point between the zero line and the linear increasing lines (dashed straight lines as examples for 5K, 20K and 50K). Similar values and temperature dependence are obtained plotting the data semilogarithmic and defining $h_{c1}^J$ at constant $\Delta m = 3 \times 10^{-7}$ emu. The red arrow shows this field for the curve measured at 20K. Inset: Temperature dependence of the lower (close squares) and upper (open circles) critical Josephson fields (left y-axis). The measured values of the lower critical field were multiplied by a factor of three in this figure. The right y-axis corresponds to the width Δm of the hysteresis field loop at zero field after cycling the field to $H_{max} = 500$ Oe. The blue curve follows $\Delta m(T) = 11.5$ [μemu] $(1 - (T/1050[K])^{1.5})$.

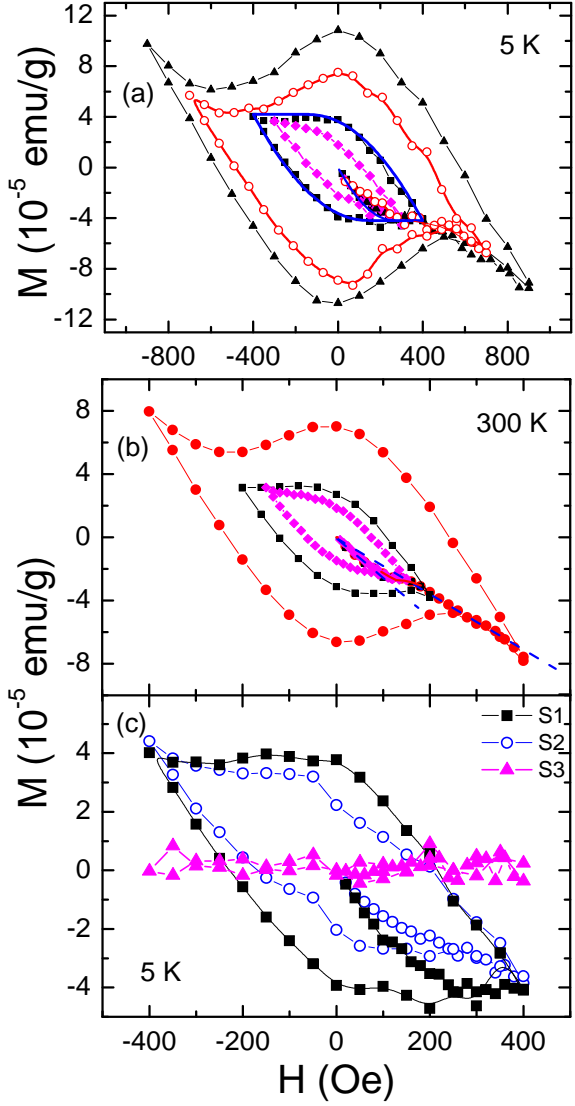

**Figure 4.** (a) Magnetization field hysteresis loops at different maximum fields $H_{max}$ (300 Oe, 400 Oe, 700 Oe and 900 Oe) and at 5K. The magnetization was calculated dividing the magnetic moment by the total sample mass. The blue continuous curve that fits the loop measured for $H_{max} = 400$ Oe was calculated with the Bean model[22] with a penetration field $H_p = 270$ Oe. A linear diamagnetic background - 6.8 x $10^{-6}$ H (Oe) emu /g was subtracted from the data. (b) The same as (a) but at 300K and for $H_{max} = 150$ Oe, 200 Oe and 400 Oe. The two dashed lines are a guide to the eye that highlight the change of slope when the field is increased from $H \ll h_{c2}^J$ to $H > h_{c2}^J$. A linear diamagnetic background - 4.6 x $10^{-6}$ H(Oe) emu/g was subtracted from the data. (c) Hysteresis loops at 5K and for $H_{max} = 400$ Oe for the water treated powder (S1, mass = 61.5 mg), the same powder but after pressing it in a pellet with a pressure of $18 \pm 5$ MPa (S2, mass = 33.6 mg) and after pressing it with a pressure of $60 \pm 20$ MPa (S3, mass = 33.7 mg). The corresponding diamagnetic linear backgrounds were subtracted from the measured data. As shown in Fig.1 for the untreated powder (as well as in Ref.[2]) the raw data has the sum of the two contributions, the diamagnetic plus that one from the superconducting regions. For example, from the results in (a) we have a ratio between the remanence to the non-superconducting diamagnetic contribution at 100 Oe of ~ 0.15 and ~ 0.015 at $10^3$ Oe. Note the difference in the field x-scale between (a) and (b,c).

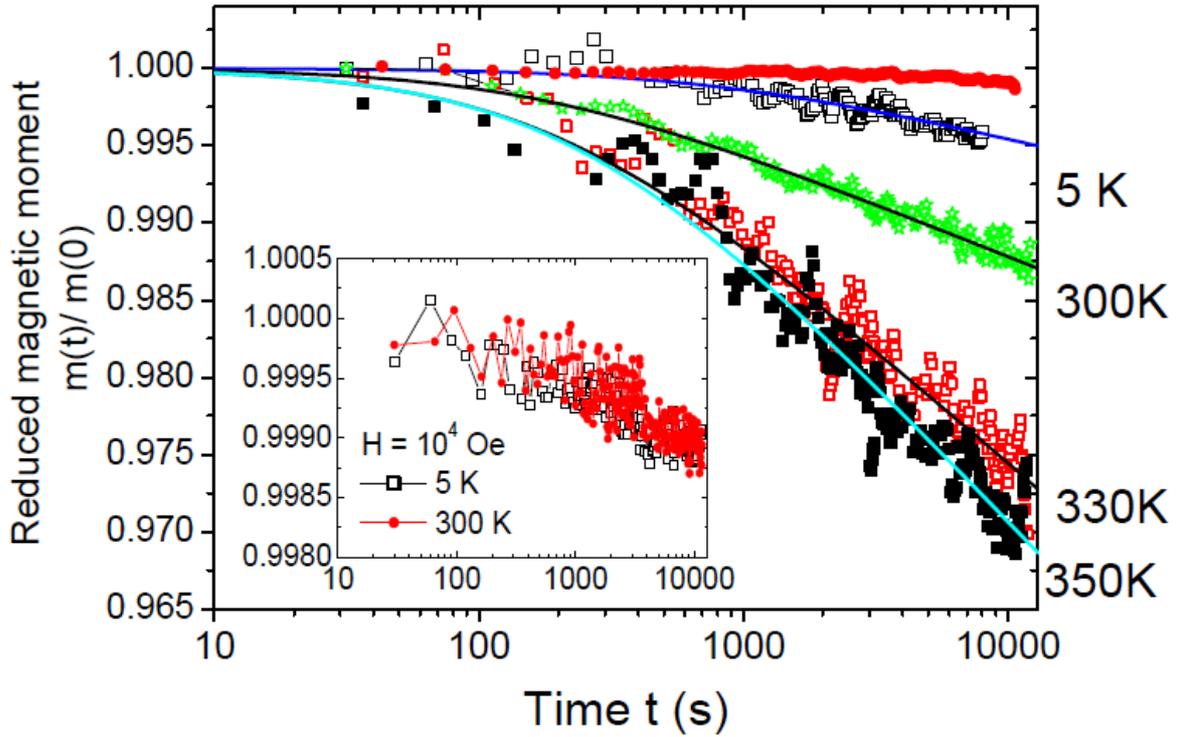

**Figure 5.** Time dependence of the normalized remanent ($H = 0$) magnetic moment at different fixed temperatures of a water treated powder of mass = 61.5 mg. Initially, a field of 300 Oe was applied to the sample and afterwards it was decreased to nominally zero field. The continuous lines are fits to the function $m(t)/m(0) = 1 - m_1 \ln(1 + (t/\tau_0))$ with $m_1$ and $\tau_0$ free parameters (for $T$ = 5K, 300K, 330K and 350K, the fitting parameters and their confidence errors are $(m_1, \tau_0)$: $(1.8 \pm 0.6) \, 10^{-3}$, $(400 \pm 200)$ s; $(3 \pm 0.5) \, 10^{-3}$, $(170 \pm 80)$ s; $(7 \pm 0.5) \, 10^{-3}$, $(250 \pm 80)$ s; $(9 \pm 0.5) \, 10^{-3}$, $(300 \pm 100)$ s. To check for possible SQUID-related artifacts in the time relaxation the red close circles (upper curve) correspond to the results obtained after a similar procedure and at 300K for a $Dy_2O_3$ sample, which had a similar absolute magnetic moment as the graphite powder sample. Inset: Similar data obtained at 5K and 300K but at a field of $10^4$ Oe coming from a state of $2 \times 10^4$ Oe.